\documentclass{emulateapj}
\usepackage{natbib}
\usepackage{bm}
\newcommand{\ra}{\;\raise1.0pt\hbox{$'$}\hskip-6pt\partial\;}
\newcommand{\lo}{\;\overline{\raise1.0pt\hbox{$'$}\hskip-6pt
\partial}\;}
\newcommand{\beq}{\begin{equation}}
\newcommand{\eeq}{\end{equation}}
\newcommand{\rhat}{\hat{\bf r}}
\begin{document}

\shorttitle{Mosaicking with CMB interferometers}
\shortauthors{Bunn \& White}

\title{Mosaicking with cosmic microwave background interferometers}
\author{Emory F. Bunn}
\affil{Physics Department, University of Richmond,
Richmond, VA 23173}
\author{Martin White}
\affil{Departments of Physics and Astronomy, University of California,
Berkeley, CA 94720}
\begin{abstract}
Measurements of cosmic microwave background (CMB) anisotropies by
interferometers offer several advantages over single-dish observations.
The formalism for analyzing interferometer CMB data is well developed
in the flat-sky approximation, valid for small fields of view.  As the
area of sky is increased to obtain finer spectral resolution, this
approximation needs to be relaxed.  We extend the formalism for CMB
interferometry, including both temperature and polarization, to 
mosaics of observations covering arbitrarily large areas of the sky,
with each individual pointing lying within the flat-sky approximation.
We present a method for computing the correlation between visibilities
with arbitrary pointing centers and baselines and illustrate
the effects of sky curvature on the $\ell$-space resolution that can
be obtained from a mosaic.
\end{abstract}

\keywords{cosmic microwave background --- techniques: interferometric}

\section{Introduction}

The study of anisotropies in the cosmic microwave background (CMB) radiation
has revolutionized cosmology.  Key to this revolution have been coupled
advances in theory, data analysis, and instrumentation.  In particular,
the design of experiments with exquisite systematic error control has been
crucial for progress in the field.
Interferometers offer several advantages in this respect, with simple
optics, instantaneous differencing of sky signals without scanning and
no differencing of detectors.  The shape of the beam can be well understood
and the measurement is done directly in Fourier space where the theory most
naturally lives.

Pioneering attempts to detect CMB anisotropy with interferometers
were made by \cite{MarPar} and \cite{Sub}.
Several groups have successfully detected primary CMB anisotropies 
\citep{CAT1,CAT2,DASIT,CBIT,VSA}
and polarization
\citep{CBIP,DASIP},
using interferometers.  The formalism for analyzing CMB data from
interferometers has been developed by
\cite{HobLasJon,HobMag,WCDH,HobMas}; and \cite{Mye};
as well as in the experimental papers cited above.
\cite{Parketal} and \cite{ParkNg} examined interferometric polarimetry.

In the Fraunhofer limit an interferometer measures the Fourier transform
of the sky, multiplied by the primary beam.  The primary beam determines
the instantaneous field of view of the instrument and its Fourier transform
is simply the autocorrelation of the Fourier transform of the point response
of the receiver to an electric field.  The angular scale probed by any pair
of telescopes being correlated is determined by their spacing in units of the
observational wavelength\footnote{We assume throughout monochromatic
radiation; the generalization to a specified frequency band is
straightforward.}.
The range of scales probed by the interferometer is then determined by the
spacing of the elements, while the resolution in spatial wavenumber is
determined by the area of sky surveyed.  By ``mosaicking'' several smaller
patches together, the resolution in spatial wavenumber can be increased,
although the range of spatial scales remains fixed by the geometry of the
interferometer elements.

In most cases it has been assumed that the field of view is small, so that
one can use the ``small-angle'' or ``flat-sky'' approximation.  However, if
we want fine resolution in spatial wavenumber -- which future experiments
are driving towards -- we need to survey large
areas of sky \citep{HobMag} and thus relax this assumption.
The purpose of this paper is to extend the formalism presented in the above
papers to the case where each individual pointing of the interferometer is 
still within the flat-sky approximation but by mosaicking many pointings
together a significant area of sky is surveyed.  Our extension allows one
to see how large an error is being made in assuming the flat-sky approximation
and shows how corrections can be systematically incorporated.

The central idea of this paper is the following.  The key ingredient
in analyzing a mosaic of interferometer pointings is the set of two-point
visibility correlations.  For each pair of pointings, we can calculate
the correlations in a spherical coordinate system that places
both pointing centers on the equator.  If each pointing has a small
field of view, then we can approximate the sphere by a cylinder
in the vicinity
of the equator, allowing the use of Fourier analysis rather than
a more cumbersome expansion in spherical harmonics.

The outline of this paper is as follows.  We begin in \S\ref{sec:flat} by
reminding the reader of some basic results in the flat-sky limit.  We
then show how this can be extended using a cylindrical projection in
\S\ref{sec:cylinder} and make contact with the exact spherical harmonic
treatment in \S\ref{sec:harmonic}.  Section \ref{sec:poln} extends our
results to include polarization, and we conclude in \S\ref{sec:conclusions}.

\section{Flat-sky limit}  \label{sec:flat}

We begin by considering the flat-sky limit and focusing on temperature
anisotropies.  Thus our interferometer is measuring a scalar field,
$T({\bf x})$, defined on the 2D plane.
In this limit the fundamental observable, a visibility, can be written
\begin{equation}
  V({\bf u}) = {\partial B_\nu\over\partial T}
    \int d^2x\ \Delta T({\bf x})\,A({\bf x})\,e^{2\pi i{\bf u}\cdot{\bf x}},
\label{eq:visdef}
\end{equation}
where $\partial B_\nu/\partial T$ converts from temperature to intensity units
and $A({\bf x})$ is the primary beam (typically normalized to unity at peak).
{}From now on we will neglect the flux-temperature conversion factors
and write $T$ for $\Delta T$.

For Gaussian fluctuations, such as the primary CMB anisotropies, we need
to compute the visibility correlation matrix
\begin{equation}
  {\cal V}_{ij} \equiv \left\langle V({\bf u}_i)V^\star({\bf u}_j)\right\rangle,
\end{equation}
where ${\bf u}_i$ and ${\bf u}_j$ represent the baselines to be correlated
and $\langle\cdot\rangle$ represents an ensemble average.  This can be related
to the usual temperature correlation function
\begin{equation}
  \left\langle T({\bf n}_i)T({\bf n}_j)\right\rangle
  = {1\over 4\pi}\sum_{\ell=2}^{\infty} (2\ell+1)C_\ell
    P_\ell({\bf n}_i\cdot{\bf n}_j)
\end{equation}
for temperatures measured in directions ${\bf n}_i$ and ${\bf n}_j$ where
$C_\ell$ are the multipole moments.  In our flat-sky limit, for a single
patch \citep[e.g.,][]{WCDH}
\begin{equation}
{\cal V}_{ij}\propto\int d^2w\,S(w)\widetilde{A}^*(2\pi[{\bf w}+{\bf u}_i])
\widetilde{A}(2\pi[{\bf w}+{\bf u}_j]),
\end{equation}
where the angular
power spectrum $S(u)$ is defined by
\begin{equation}
  (2\pi)^2\ u^2S(u) \simeq \ell(\ell+1)C_\ell \quad {\rm for}\ \ell=2\pi u,
\label{eq:S}
\end{equation}
and $\widetilde A$ is the Fourier transform of the antenna pattern,
\begin{equation}
\widetilde A({\bf k})=(2\pi)^{-2}\int d^2x \,A({\bf x})e^{-i{\bf k}\cdot{\bf x}}.
\end{equation}
The extension to multiple different patches, each with its own pointing
center ${\cal P}$, merely inserts a phase factor
\begin{equation}
  {\cal V}_{ij}^{{\cal P}_1{\cal P}_2} \propto \int d^2w \,S(w)
  \widetilde{A}^*(2\pi[{\bf w}+{\bf u}_i])\widetilde{A}(2\pi[{\bf w}+{\bf u}_j])
  e^{2\pi i{\bf w}\cdot{\bf D}},
\label{eq:v12flat}
\end{equation}
where ${\bf D}$ is the separation between the pointing centers
${\cal P}_1$ and ${\cal P}_2$.  From now on we will drop the superscripts
on ${\cal V}_{ij}$.

\section{Cylindrical method} \label{sec:cylinder}

The flat-sky approximation above is valid only if both the field of view
of an individual pointing and the separation ${\bf D}$ between pointings
are small.  We will now assume a mosaic of pointings that
cover a large area, although each individual pointing observes
only a small patch of sky.
We therefore relax the second assumption while retaining the first.

For a statistically isotropic temperature field we are free to use any
coordinate system we like to compute the visibility correlation
${\cal V}_{12}$.  In particular, we can
arrange to have the two pointing centers lie on the equator of a spherical
coordinate system ($\theta=\pi/2$) and be separated by an angle $\beta$.
We introduce a cylindrical coordinate system, with the cylinder tangent to the
sphere at the equator, denoted by $\bm{\xi}=(\phi,z)$.  Since both observations
sample only regions near the equator, we can pretend that the data
live on the cylinder rather than on the sphere.  In this approximation, 
it is natural
to expand the temperature $T({\bf x})$ in a discrete Fourier series in $\phi$
and a continuous Fourier transform in $z$:
\begin{equation}
  T(\bm{\xi}) = \sum_m\int dn\ \widetilde{T}_m(n) e^{i(m\phi+nz)}
\end{equation}
with
\begin{equation}
\left\langle \widetilde{T}_m(n)\widetilde{T}_{m'}^\star(n')\right\rangle
  = 
{{\cal P}(\sqrt{m^2+n^2})\over (2\pi)^2}
\delta_{mm'}\delta(n-n').
\end{equation}
The power spectrum is ${\cal P}(k)=(2\pi)^2S(k/2\pi)$ with
$S$ as defined in Eq.~(\ref{eq:S}).   It is related to the
spherical harmonic angular power spectrum  by ${\cal P}(k)\simeq C_k$ for
large $k$.

The visibility becomes
\begin{equation}
  V({\cal P}_1,{\bf u}_1) = (2\pi)^2\int dn\sum_m
  \widetilde{T}_m(n)\widetilde{A}^\star\left(2\pi[{\bf u}_1+{\bf w}]\right),
\label{eq:vis}
\end{equation}
where the vector ${\bf w}$ has coordinates $(w_\phi,w_z)=(m,n)/(2\pi)$ and
\begin{equation}
\widetilde{A}({\bf k})=
  \int {d^2\xi\over(2\pi)^2}\, A(\bm{\xi})e^{-i{\bf k}\cdot\bm{\xi}}
\end{equation}
is the usual Fourier transform of the primary beam.  Since we imagine
$A$ is non-zero only over a small region we can extend the integral over
the entire plane.

\begin{figure}
\begin{center}
\resizebox{!}{2.5in}{\includegraphics{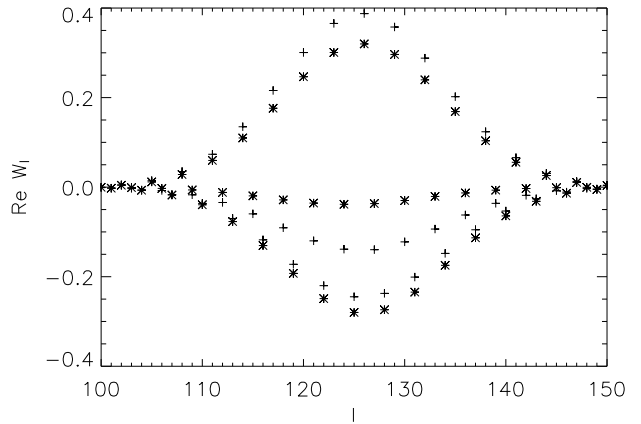}}
\end{center}
\caption{
Window functions for the covariance between two interferometer pointings
in both the flat and cylindrical approximations.  The antenna
pattern is Gaussian with beam width $\sigma=5^\circ$.  The two pointings
are separated by $120^\circ$.  Each visibility has a baseline
of magnitude $u=20$ pointing in the $\hat{\bm\phi}$ direction.  The plus signs 
are the cylindrical approximation, and the stars are the flat approximation.
For clarity, only the real parts of the window functions are shown.
\vskip 0.2in
}
\label{fig:flatvscyl}
\end{figure}

The visibility for the second pointing center is analogous, except for
a phase factor $e^{im\beta}$, so the correlation between two pointings
becomes
\begin{equation}
  {\cal V}_{12} \propto 
  \int dn\sum_m\ S(w) {\cal W}_{12}({\bf u}_1,{\bf u}_2,{\bf w}) e^{im\beta},
\label{eqn:v12cylinder}
\end{equation}
where we have defined the window function
\begin{equation}
  {\cal W}_{12}({\bf u}_1,{\bf u}_2,{\bf w}) \equiv
  \widetilde{A}^\star\left(2\pi[{\bf u}_1+{\bf w}]\right)
  \widetilde{A}      \left(2\pi[{\bf u}_2+{\bf w}]\right)
\label{eq:window}
\end{equation}
It is convenient to define a window function that is averaged over direction:
\begin{equation}
{\cal V}_{12}=\sum_\ell W^{(12)}_\ell C_\ell.
\end{equation}
To compute $W_\ell^{(12)}$, we divide 
the integral and sum in equation (\ref{eqn:v12cylinder})
into bands with $\ell-{1\over 2}<2\pi w<\ell +{1\over 2}$.  Within
each band we assume the power spectrum is constant and remove it from
the integral.  We can calculate this window function in the flat-sky
approximation instead of the cylindrical approximation if we like,
simply by replacing the sum over $m$ by an integral.

Figure \ref{fig:flatvscyl} illustrates the difference between the
flat and cylindrical approximations for large $\beta$.  The difference
is most significant when the baseline vectors ${\bf u}_i$ are nearly 
equal in magnitude and parallel
to the separation direction $\hat{\bm\phi}$: otherwise the correlation
${\cal V}_{12}$ is always small whenever $\beta$ is large.

We can use this prescription to calculate the full visibility covariance
matrix for a mosaic of many pointings.  For each pair of pointings, we must
transform to a coordinate system in which both pointings lie on the equator.
In performing this rotation, the components of the baseline vectors ${\bf u}_i$
will naturally be transformed.  The Appendix contains an explicit recipe
for performing this rotation.

\begin{figure}
\begin{center}
\resizebox{!}{2.5in}{\includegraphics{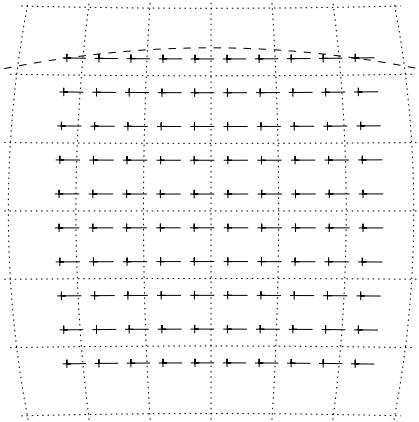}}
\end{center}
\caption{Pointing centers ($+$) and baseline vectors (horizontal
bars) for the mosaicking
example described at the end of section \ref{sec:cylinder}.  In
calculating the covariance between the two visibilities at the upper
corners of the grid, we must rotate to a coordinate system in which
the great circle connecting them (the dashed line) becomes the equator.
Note that the two baseline vectors are not parallel in the new coordinate
system.
\vskip 0.1in
}
\label{fig:grid}
\end{figure}

\begin{figure}
\begin{center}
\resizebox{!}{2.5in}{\includegraphics{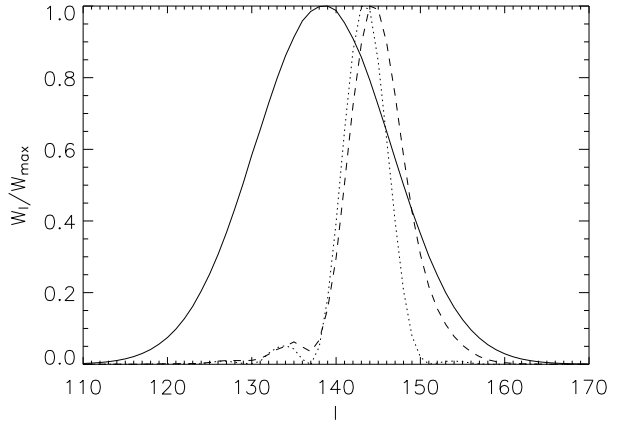}}
\end{center}
\caption{
Improvement in $\ell$-space resolution due to mosaicking.
The solid curve is the window function for a single pointing.
The dashed curve is the window function for the sum of all pointings
in a $10\times 10$ grid, after correctly accounting for baseline
rotation.  The dotted curve is the window function that would be obtained
by incorrectly assuming a flat sky over the entire mosaic.  All three
window
functions have arbitrary normalization.
\vskip 0.2in}
\label{fig:mosaicres}
\end{figure}

Figures \ref{fig:grid} and \ref{fig:mosaicres} present a simple illustration
of how mosaicking increases the $\ell$-space resolution of an experiment.  
In each pointing, the beam pattern is a Gaussian with beam width
$\sigma=5^\circ$.  We assume a $10\times 10$ grid of pointings, separated
by $5^\circ$ in both $\theta$ and $\phi$ in a spherical coordinate system,
with the center of the grid on the equator ($\theta=\pi/2$).  For each
pointing, we consider only a single baseline ${\bf u}=22\hat{\bm\phi}$.
Figure \ref{fig:grid} shows the locations of the pointing centers
and the baselines in Aitoff projection.
Note that although all the baselines have identical components in the
spherical coordinate system, they do not when rotated to the appropriate
coordinate system for computing the covariances.
As an example, to compute the covariance between the two pointings
in the upper corners of the grid, we must use a coordinate system
in which the great circle represented by the dashed line becomes the
equator.  In this coordinate system, the two baseline vectors
have $\hat{\bm\theta}$ components of opposite sign.

The visibility obtained from any single pointing provides an estimate of
the power spectrum  with a fairly wide window function (solid curve in
Fig.~\ref{fig:mosaicres}).
We can obtain an estimate
of the power spectrum with a narrower window function by
simply adding together all 100 visibilities.  To find the
window function for the sum, we write down the absolute
square of the sum of all 100 visibilities:
\beq
\left<\left|\sum_{i=1}^{100}V_i\right|^2\right>
=\sum_{i,j=1}^{100}\langle V_iV_j^*\rangle
=\sum_{i,j=1}^{100}{\cal V}_{ij}.
\eeq
The window function for the sum of all the visibilities is therefore
the sum of $W_l^{(ij)}$ over all visibility pairs $i,j$.  We can
compute each $W_l^{(ij)}$ using the recipe described above.
Specifically, for each pair of pointings, we find
a rotation 
that brings both pointing centers to the equator, apply that rotation
to the vectors ${\bf u}_1,{\bf u}_2$ using the recipe in the Appendix, 
and apply equation (\ref{eq:window}).

The dashed curve in Figure \ref{fig:mosaicres} is the window function
for the sum of all pointings.  As expected, the $\ell$-space resolution
has improved.
The dotted curve shows the window function obtained by incorrectly
assuming the sky is flat over the entire mosaic -- that is, assuming
that all of the baseline vectors illustrated in Fig.~\ref{fig:grid}
lie in the same plane and are parallel.

The difference between the dashed and dotted curves
is almost entirely due to  ``baseline rotation'' -- the fact that, e.g., the 
baseline vectors at the corners of Fig.~\ref{fig:grid} are not in fact
parallel to each other when viewed in a coordinate system in which
both lie on the equator.
It makes virtually
no difference whether we use the 
flat method (integral over $m$), the cylindrical method (sum over $m$),
or an exact spherical harmonic calculation as described below
in calculating each $W_\ell^{(ij)}$, as long as we get the baseline
rotation right.

Of course, other linear combinations could be used instead
of a simple sum of all 100 pointings, resulting in window functions
with peaks in different places (within the envelope set by the
single-pointing window function).

\section{Harmonic method} \label{sec:harmonic}

We can also make direct contact with usual spherical harmonic treatment
in which
\begin{equation}
  T(\hat{\bf r}) = \sum_{\ell m} a_{\ell m}Y_{\ell m}(\hat{\bf r})
\label{eqn:harmonic}
\end{equation}
and
\begin{equation}
  \left\langle a_{\ell m}a_{\ell' m'}^\star\right\rangle
  = C_\ell \delta_{\ell\ell'}\delta_{mm'} .
\end{equation}
The visibility for a single pointing is 
\begin{equation}
V({\bf u})=\int d^2\rhat\,A(\rhat)T(\rhat)e^{2\pi i{\bf u}\cdot\rhat}=
\sum_{\ell,m}a_{\ell m}F_{\ell m}({\bf u}),
\label{eqn:visexact}
\end{equation}
where
\begin{equation}
F_{\ell m}({\bf u})=\int A(\rhat)Y_{lm}(\rhat)e^{2\pi i{\bf u}\cdot\rhat}.
\end{equation}
It is of course possible to perform these integrals numerically
and calculate the visibility covariance matrix without any approximations
at all.  In this section we will see how to obtain the cylindrical
approximation from this exact expression.

In previous treatments \citep[e.g.,][]{WCDH}, the flat-sky
limit for a single pointing 
was taken by approximating the spherical harmonics near the pole of
the spherical coordinate system ($\theta=0$).  To obtain the visibility
covariance for two different pointings it is more convenient
to place the pointing centers on the equator as in the previous section.

Near the equator $z\equiv\cos\theta\simeq\pi/2-\theta$.
Using the recurrence relations for the associated Legendre polynomials
\citep{AbrSte,GraRyz},
one can show
\begin{equation}
  Y_{\ell m}(\phi,z) \to N_{\ell m} e^{im\phi} \left\{
  \begin{array}{ll}
  \hphantom{-i}\cos n_{\ell m}z & {\rm if}\ \ell-m\ {\rm even} \\
  -i\sin n_{\ell m}z & {\rm if}\ \ell-m\ {\rm odd}
  \end{array}\right.
\label{eq:ylmplane}
\end{equation}
with
\begin{equation}
  n_{\ell m} = \sqrt{ \ell(\ell+1)-m^2 }
\end{equation}
and
\begin{eqnarray}
  N_{\ell m} &=& (-1)^{(\ell+m)/2}\ {2^{m}\over\sqrt{\pi}}
  \sqrt{{2\ell+1\over 4\pi}} \nonumber \\
  &\times&
  \sqrt{{(\ell-m)!\over(\ell+m)!}}\ 
  {\left([\ell+m-1]/2\right)!\over\left([\ell-m]/2\right)!},
\end{eqnarray}
which can also be written
\begin{eqnarray}
  N_{\ell m} &=& (-1)^{(\ell+m)/2}\ 2^{-\ell} \sqrt{{2\ell+1\over 4\pi}}
  \nonumber \\
  &\times&
  {\sqrt{(\ell+m)!(\ell-m)!}\over
   \left({\ell+m\over 2}\right)!\left({\ell-m\over 2}\right)!}
\end{eqnarray}
by using \citep{GraRyz}
\begin{equation}
  \left( n+{1\over 2}\right)! = \sqrt{\pi}\ {(2n+1)!\over 2^{2n+1}\,n!} .
\end{equation}
In all of these expressions the factorials should be interpreted as $\Gamma$
functions for non-integer arguments.  In the limit when all of the
factorial moments are large we can use the approximation
\citep{AbrSte}
\begin{equation}
  \ln N! \simeq \left(N+{1\over 2}\right)\ln N - N + {\rm const}
\end{equation}
to write the normalization factor as
\begin{equation}
  N_{\ell m}\simeq \frac{(-1)^{(l+m)/2}}{\pi}
  \frac{\sqrt{l+\frac{1}{2}}}{(l^2-m^2)^{1/4}}
\end{equation}

Note that the $Y_{\ell m}$ are eigenfunctions of the 2D Laplacian with
eigenvalues $-\ell(\ell+1)$, and the form of $n_{\ell m}$ guarantees
that this is preserved in the cylindrical coordinates:
\begin{eqnarray}
  \nabla^2Y_{\ell m} &\to& \left({\partial^2\over\partial\phi^2}+
  {\partial^2\over\partial z^2}\right)Y_{\ell m} \\
  &=& \left( -m^2 - n_{\ell m}^2 \right)Y_{\ell m} \\
  &=& -\ell(\ell+1)Y_{\ell m},
\end{eqnarray}
where the arrow indicates the cylindrical coordinate limit.

If we define
\begin{equation}
  \alpha_{\ell m} = (-1)^{\ell+m} {N_{\ell m}\over 2} a_{\ell m},
  \qquad
  \bar{\alpha}_{\ell m} = {N_{\ell m}\over 2} a_{\ell m},
\end{equation}
then we can rewrite Eq.~(\ref{eqn:harmonic}) as
\begin{equation}
  T(\bm{\xi}) = \sum_{\ell m} \alpha_{\ell m}e^{i{\bf k}\cdot\bm{\xi}}
  + \bar{\alpha}_{\ell m}e^{i\bar{{\bf k}}\cdot\bm{\xi}}
\end{equation}
with the definitions ${\bf k}=(m,n_{\ell m})$ and
$\bar{{\bf k}}=(m,-n_{\ell m})$.  From now on we will take the sum over
both $\pm n_{\ell m}$ as implicit and write
\begin{equation}
  T(\bm{\xi}) = \sum_{{\bf k}} \alpha_{\bf k}e^{i{\bf k}\cdot\bm{\xi}}.
\end{equation}
This way of writing the spherical harmonic expansion
makes the correspondence with the Fourier representation explicit.

The visibility becomes
\begin{equation}
  V({\bf u}) = (2\pi)^2 \sum_{\bf k} \alpha({\bf k})
    \widetilde{A}^\star\left({\bf k}+2\pi{\bf u}\right),
\end{equation}
and the correlation matrix is
\begin{equation}
  {\cal V}_{12} = {(2\pi)^4\over 4}\sum_{\bf k} \left|N_{\ell m}\right|^2 C_\ell
  {\cal W}_{12}({\bf u}_1,{\bf u}_2,{\bf k}/2\pi)e^{i{\bf k}\cdot\beta},
\label{eqn:v12harmonic}
\end{equation}
plus oscillatory terms that average to zero in the sum over $\ell$ and $m$.

If we work at large values of $\ell$ and $m$ we can replace the sum over
${\bf k}$ with $\sum_m$ and $\int dn$ and recover our cylindrical result,
Eq.~(\ref{eqn:v12cylinder}).  One can verify by explicit computation that
using
\begin{equation}
  {d\ell\over dn} = {\sqrt{\ell(\ell+1)-m^2}\over\ell+{1\over 2}}
\end{equation}
to turn the sum over $\ell$ into an integral
and using the asymptotic form of $N_{\ell m}$ in
Eq.~(\ref{eqn:v12harmonic}) leads to Eq.~(\ref{eqn:v12cylinder}).

\begin{figure}
\begin{center}
\resizebox{!}{3in}{\includegraphics{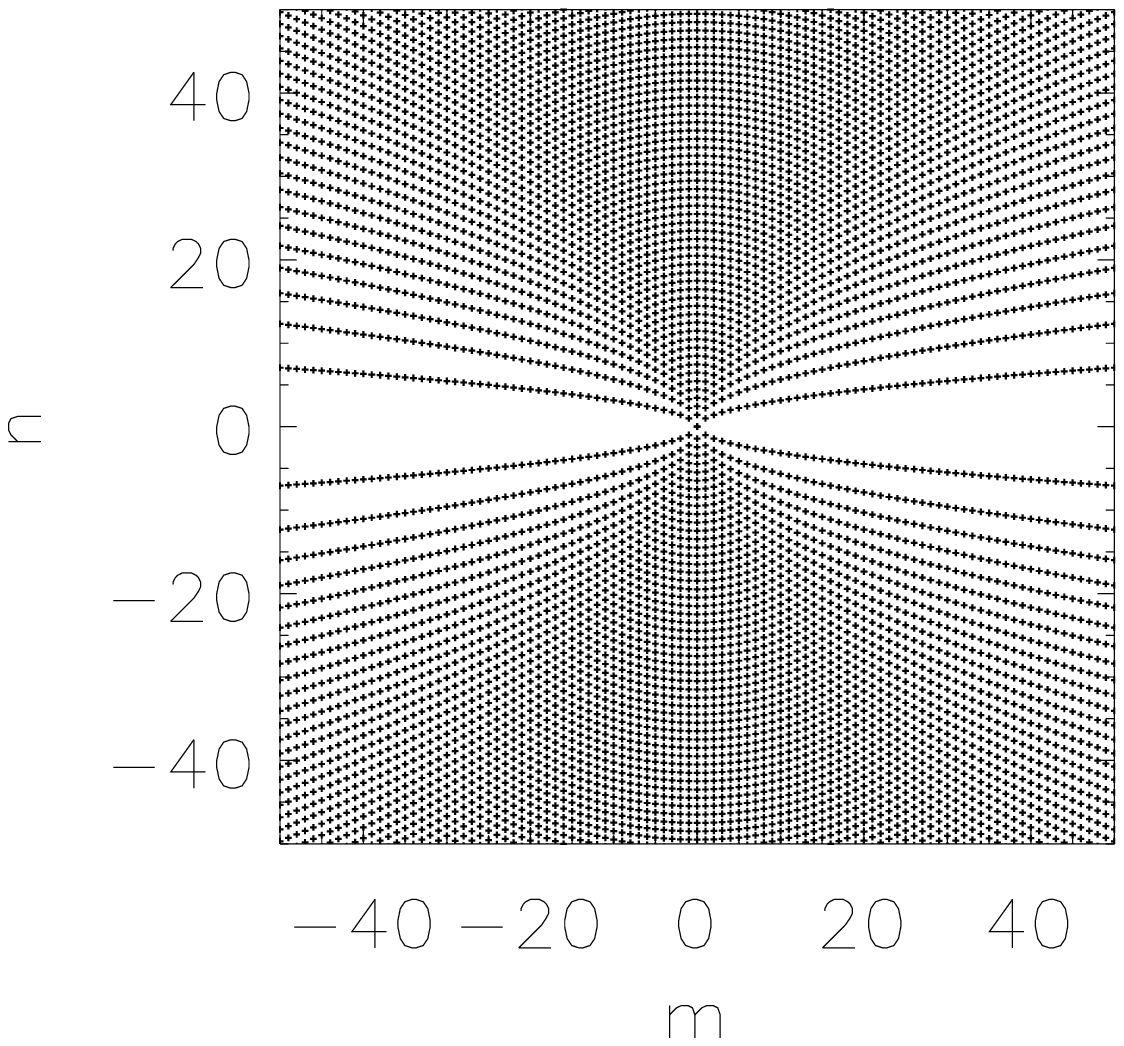}}
\end{center}
\caption{The Fourier modes ${\bf k}_{lm}^\pm=(m,\pm n_{lm})$.
\vskip 0.2in}
\label{fig:kspace}
\end{figure}

As Fig.~\ref{fig:kspace} illustrates, 
the differences between the harmonic and Fourier expansions lie in the
non-uniform gridding of the wave vectors ${\bf k}$.  For any given
wavenumber $\ell$ the modes are packed more sparsely near the $m$
axis, corresponding to $|m|\simeq\ell$, and more densely near $m\simeq 0$.
The normalization factor $N_{\ell m}$ weights
the higher $m$ modes more strongly to make up for this.

\begin{figure}
\begin{center}
\resizebox{!}{2.5in}{\includegraphics{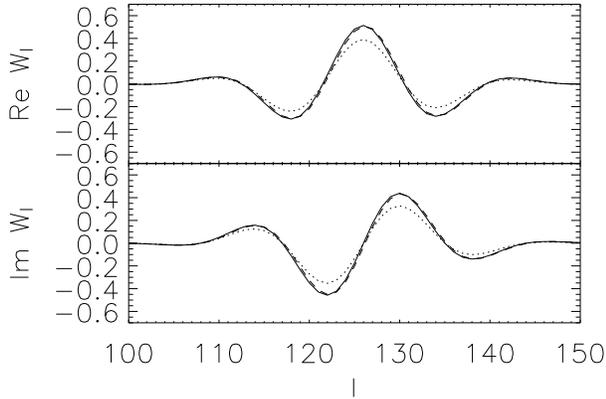}}
\end{center}
\caption{Comparison of approximations.  The window function is shown
for the covariance between visibilities with ${\bf u}=20\hat{\bm\phi}$, beam
width $\sigma=5^\circ$, and pointing centers separated by $\beta=15^\circ$.
The solid curve is the exact spherical harmonic calculation; the dashed
curve is the cylindrical approximation; and the dotted curve is the 
approximation obtained by approximating the spherical harmonics by
plane waves.
\vskip 0.2in}
\label{fig:approximations}
\end{figure}

Surprisingly, in some cases
the cylindrical approximation proves numerically more accurate than
the approximation in this section, as illustrated in 
Fig.~\ref{fig:approximations}.  Specifically, in cases where the baseline
vectors point in the $\hat{\bm\phi}$ direction, the sum in equation 
(\ref{eqn:v12harmonic}) is dominated by modes with $|m|=\ell$, which are 
precisely the modes for which the plane-wave approximation to $Y_{\ell m}$
is worst.  We found no instance in which the cylindrical approximation 
(\ref{eq:ylmplane}) does
worse than the approximate spherical harmonic expansion of this section,
so for numerical work one should either use the cylindrical approximation
or the exact full-sky expressions.  In general, we find that
the cylindrical approximation
starts to become poor for Gaussian beam widths $\sigma\simeq 8^\circ$ 
(FWHM $\simeq 20^\circ$).

\section{Polarization} \label{sec:poln}

In this section, we will generalize the results of the previous sections
to observations of linear polarization.  Instead of considering a single
scalar observable $T$, we must consider observations of the two
Stokes parameters $Q,U$.

It is convenient to combine the Stokes parameters into the complex
quantities
\beq
P_\pm={1\over\sqrt{2}}(Q\pm iU),
\eeq
because these quantities transform in a relatively simple way
under rotations: under a rotation by an angle $\psi$ about a given
point $\rhat$, $P_\pm(\rhat)\to P_\pm(\rhat)e^{\mp 2i\psi}$.  In
other words, $P_+$ is a quantity of spin weight $-2$ and $P_-$ has
spin weight $+2$.  These transformation properties make $(P_+,P_-)$ 
a more convenient basis of observables to work with than $(Q,U)$.  The
two bases are related by a unitary transformation,
\beq
\left(\matrix{P_+\cr P_-}\right)=
  {1\over\sqrt 2}\left(\matrix{1 & 1\cr -i & i}\right)
  \left(\matrix{Q\cr U}\right),
\label{eqn:unitary}
\eeq
so any results derived in one basis can easily be transformed to the
other.

An interferometer that works by interfering circularly
polarized radiation from the two antennas measures the visibilities
\beq
V_\pm({\bf u})=\int d^2\rhat \,P_\pm(\rhat)A(\rhat)e^{2\pi i{\bf u}\cdot\rhat}.
\eeq
Specifically, interfering left-circularly polarized radiation from antenna 1
with right-circularly polarized radiation from antenna 2 yields $V_+$, and
reversing the senses of both circular polarizations yields $V_-$.
(Interfering right with right and left with left yields visibilities that
probe total intensity and circular polarization.)
On the other hand, an interferometer that works by combining linear
polarization states would measure visibilities $V_Q$ and $V_U$ for the
individual Stokes parameters.
(For instance, interfering $E_x$ from antenna 1 with $E_y$ from antenna 2
yields $V_U$.)  
We will assume that the measured\footnote{We are ignoring some sources of
systematic error in this expression.  For instance, in an instrument with
cross-polar beam response, each measured visibility would contain contributions
{}from both $P_+$ and $P_-$, with different effective antenna patterns.}
quantities are $V_\pm$ rather than $V_{Q,U}$, but all results are easily
transformed to the $Q,U$ basis using Eq.~(\ref{eqn:unitary}).

As in the case of temperature anisotropy, the key ingredient in analyzing
CMB interferometric observations of polarization is the visibility
covariance matrix:
\beq
{\cal V}_{12}^{\pm\bm\pm}\equiv \langle V_{\pm}({\bf u}_1,{\cal P}_1)
V_{\bm\pm}({\bf u}_2,{\cal P}_2)^*\rangle,
\label{eqn:viscovarpol}
\eeq
where ${\bf u}_i$ and ${\cal P}_i$ represent baselines and pointing
centers for a pair of visibilities.  
Note that in this equation the signs of $\pm$ and $\bm\pm$ can
be varied independently --- that is, there are in general
four distinct 
covariances, ${\cal V}_{12}^{++},{\cal V}_{12}^{+-},{\cal V}_{12}^{-+},
{\cal V}_{12}^{--}$.

Our primary interest will continue to be the case where the flat-sky
approximation is appropriate for each individual pointing but the separation
between pointings is not necessarily small.  We will present 
exact expressions
for the visibility covariances in terms of spherical harmonics first,
then show that they reduce in the this limit to cylindrical-sky
expressions similar to the anisotropy results above.

\begin{figure*}
\begin{center}
\resizebox{!}{2.5in}{\includegraphics{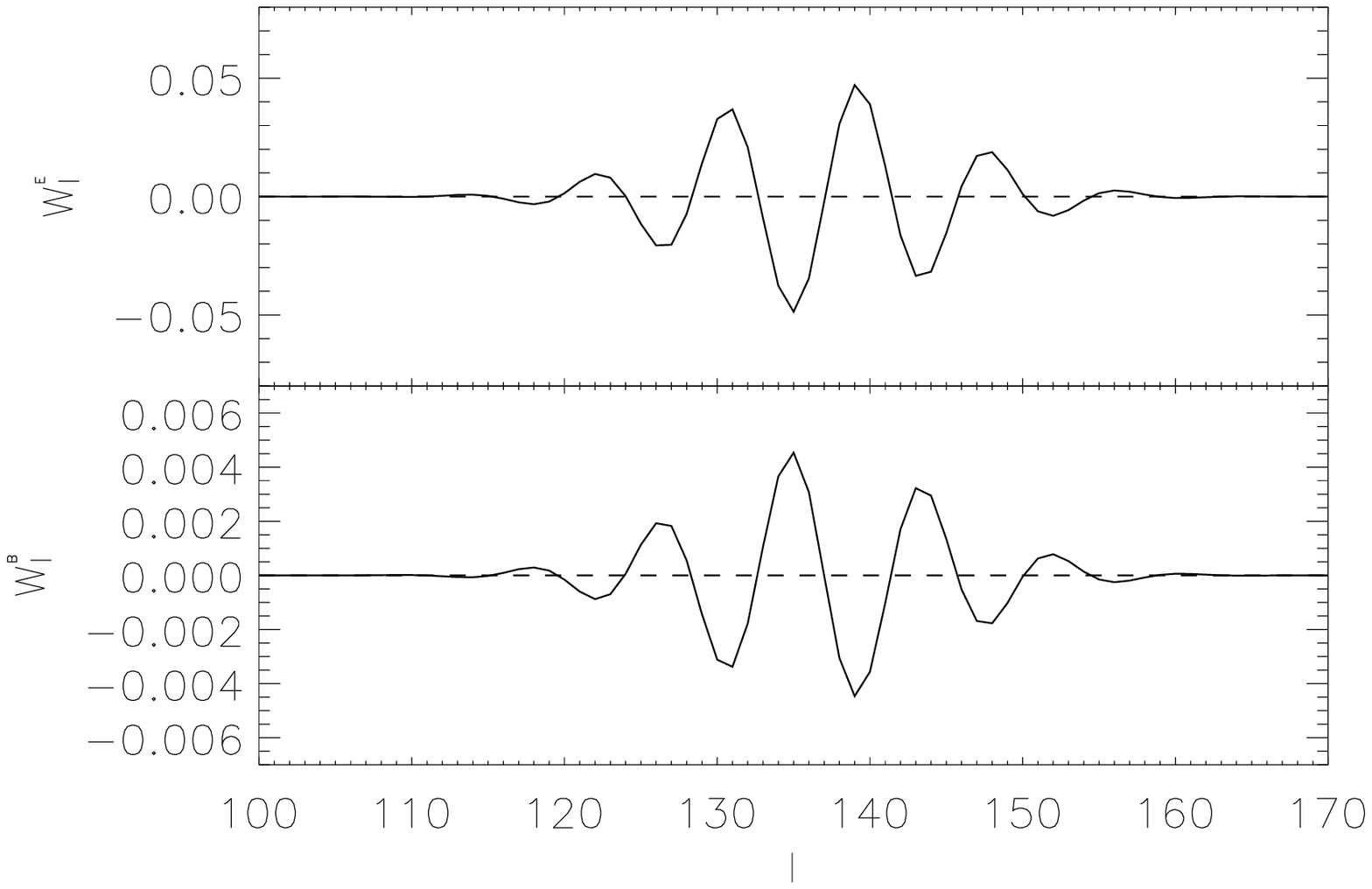}}
\resizebox{!}{2.5in}{\includegraphics{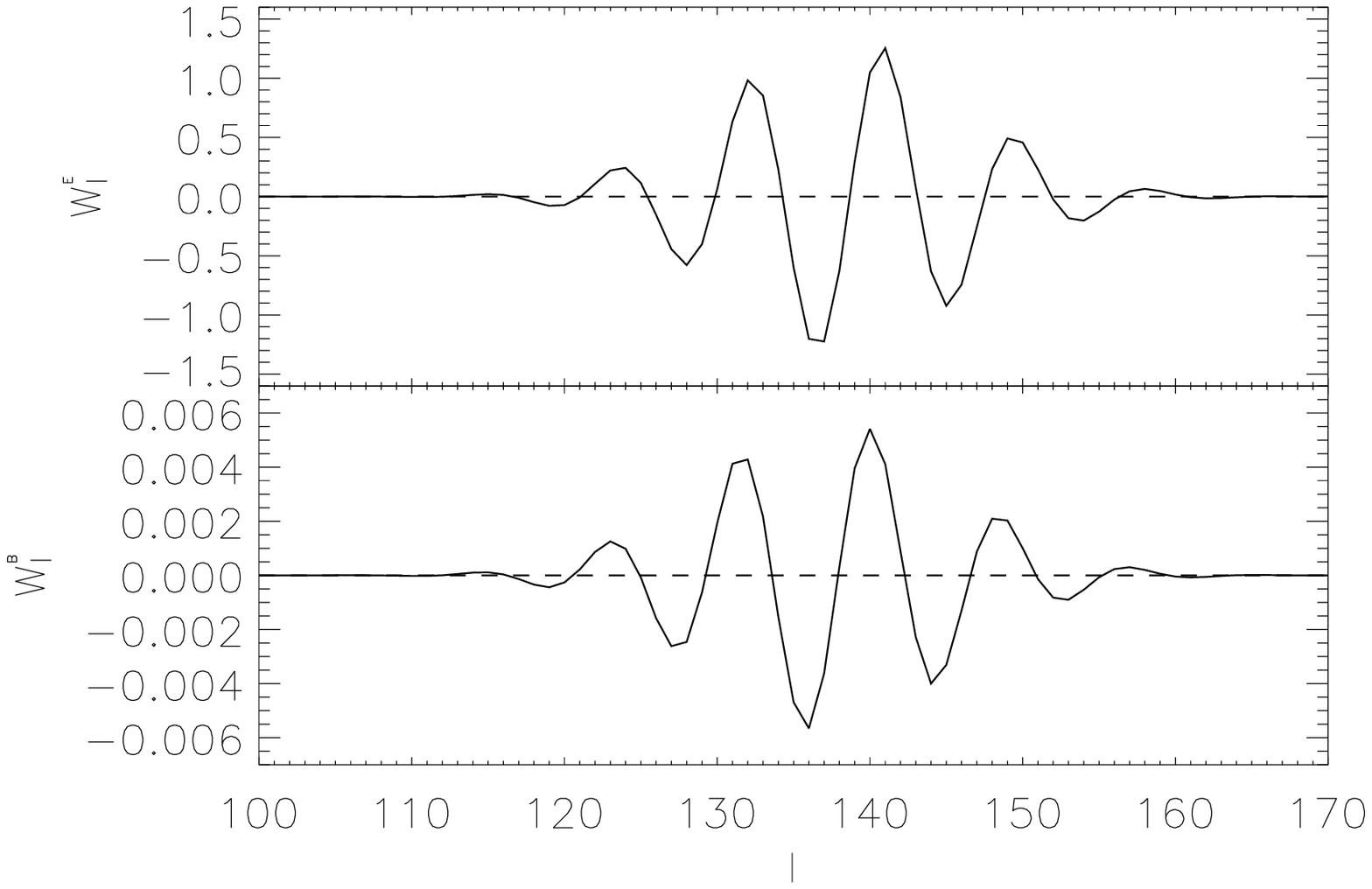}}
\end{center}
\caption{Polarization window functions.  The left panel shows
the window functions for the correlation $\langle V_QV_Q^*\rangle$
of the two baselines at the upper corners of Fig.~\ref{fig:grid}.
The right panel shows the correlations calculated by incorrectly
assuming the entire mosaic is flat -- that is, ignoring the rotation
of the basis vectors $\hat{\bm\theta},\hat{\bm\phi}$ when moving from one point to
another.  Note that the vertical axes of the upper plots differ
by a factor of 200.
These window functions were calculated in the cylindrical
approximation, but the exact spherical harmonic calculation yields
negligible differences.
\vskip 0.2in}
\label{fig:polwindows}
\end{figure*}

\subsection{Flat sky}

Assume our observations cover a small enough patch of sky
that we can replace spherical harmonic expansions with Fourier
transforms:
\beq
P_{\pm}({\bf x})=\int d^2k\,\tilde P_\pm({\bf k})e^{i{\bf k}\cdot {\bf x}}.
\eeq
The two Fourier transforms are related like this:
\beq
\tilde P_\pm^*({\bf k})=\tilde P_{\mp}(-{\bf k}).
\eeq

A key insight into the analysis of CMB polarization was the observation
\citep{kks,zalsel}
that any polarization field can be decomposed
into a scalar part (conventionally denoted $E$) and a pseudoscalar
part (denoted $B$).
The $E$-$B$ separation is particularly simple in Fourier space: modes
with polarization parallel or perpendicular to ${\bf k}$ are $E$ modes,
while modes polarized at $45^\circ$ are $B$ modes.  In terms of $\tilde P_\pm$,
this means that
\beq
\tilde P_{\pm}({\bf k})=(\tilde E({\bf k})\pm i\tilde B({\bf k}))
e^{\pm 2i\psi_{\bf k}},
\eeq
where $\psi_{\bf k}$ is the angle made by the wavevector ${\bf k}$
with respect to the $x$ axis.

Assuming that the polarization 
was generated by a homogeneous, isotropic, parity-respecting
process, the two-point correlations between
$E$ and $B$ are determined by two power spectra ${\cal P}_{E,B}$
\begin{eqnarray}
\langle\tilde E({\bf k})\tilde E^*({\bf k}')\rangle&=&(2\pi)^{-2}
{\cal P}_E(k)\delta(
{\bf k}-{\bf k'}),\\
\langle\tilde B({\bf k})\tilde B^*({\bf k}')\rangle&=&(2\pi)^{-2}{\cal P}_B(k)
\delta(
{\bf k}-{\bf k'}),\\
\langle\tilde E({\bf k})\tilde B^*({\bf k}')\rangle&=&0.
\end{eqnarray}

This means that the covariances of the polarization $\tilde P_{\pm}$ are
\begin{eqnarray}
\langle \tilde P_{\pm}({\bf k})\tilde P_{\pm}^*({\bf k}')\rangle
&=&(2\pi)^{-2}({\cal P}_E(k)+{\cal P}_B(k))\times\nonumber\\
& &\qquad\delta({\bf k}-{\bf k}'),\\
\langle \tilde P_{\pm}({\bf k})\tilde P_{\mp}^*({\bf k}')\rangle
&=&(2\pi)^{-2}({\cal P}_E(k)-{\cal P}_B(k))\times\nonumber\\
& & \qquad e^{\pm 4i\psi_{\bf k}}\delta({\bf k}-{\bf k}').
\end{eqnarray}

Just as in the scalar case, the visibility associated
with a pointing center ${\bf b}$ and a baseline ${\bf u}$
can be expressed in terms of the Fourier transform of the antenna
pattern:
\beq
V_{\pm}({\bf u},{\bf b})=(2\pi)^2\int d^2k\,\tilde A^*({\bf k}+
2\pi{\bf u})\tilde P_\pm({\bf k})e^{i{\bf k}\cdot{\bf b}}.
\eeq
The correlation between two visibilities is
\begin{eqnarray}
&&{\cal V}_{12}^{\pm\pm}
=(2\pi)^2
\int d^2k\,\tilde A^*({\bf k}+2\pi {\bf u}_1)\tilde A({\bf k}+2\pi {\bf u}_2)
\nonumber\\&&
\qquad\qquad
e^{i{\bf k}\cdot({\bf b}_1-{\bf b}_2)}({\cal P}_E(k)+{\cal P}_B(k)),\\
&&{\cal V}_{12}^{\pm\mp}=(2\pi)^2
\int d^2k\,\tilde A^*({\bf k}+2\pi {\bf u}_1)\tilde A({\bf k}+2\pi {\bf u}_2)
\nonumber\\&&\qquad\qquad
e^{i{\bf k}\cdot({\bf b}_1-{\bf b}_2)}({\cal P}_E(k)-{\cal P}_B(k))
e^{\pm 4i\psi_{\bf k}}.
\end{eqnarray}

\subsection{Spherical harmonics} \label{sec:polcylinder}

Since the quantities $P_\pm$ are quantities of spin weight $\mp 2$,
it is natural to expand them in spin-($\pm 2$) spherical harmonics:
\beq
P_{\pm}(\rhat)=\sum_{\ell,m}a_{\mp 2,\ell m}
\ {}_{\mp 2}Y_{\ell m}(\rhat).
\eeq

The decomposition into $E$ and $B$ components is particularly simple 
in terms of the spherical harmonic coefficients:
\beq
a_{\pm 2,\ell m}=E_{\ell m}\pm iB_{\ell m}.
\label{eqn:almpol}
\eeq
The two-point statistics are completely described by
two power spectra $C_\ell^{EE},C_\ell^{BB}$:
\begin{eqnarray}
\langle E_{\ell m}E_{\ell'm'}^*\rangle&=&C_\ell^{EE}\delta_{\ell\ell'}
\delta_{mm'},\\
\langle B_{\ell m}B_{\ell'm'}^*\rangle&=&C_\ell^{BB}\delta_{\ell\ell'}
\delta_{mm'},\\
\langle E_{\ell m}B_{\ell'm'}^*\rangle&=&0.
\end{eqnarray}
As in the case of temperature anisotropy, the spherical and flat-sky
power spectra are related via $C_\ell\simeq {\cal P}(u)$ with $l=2\pi u$.

The visibilities can be expressed in terms of the spherical harmonic
coefficients as
\beq
V_\pm({\bf u})=\sum_{\ell,m}a_{\mp 2,\ell m} F_{\mp 2,\ell m}({\bf u}),
\label{eqn:vispolalm}
\eeq
where
\beq
F_{\mp 2,\ell m}({\bf u})=
\int d^2\rhat\,A(\rhat)\ {}_{\mp 2}Y_{\ell m}(\rhat)
e^{2\pi i{\bf u}\cdot\rhat}.
\eeq

Consider first the covariance between two visibilities with
identical pointing centers.  Combining equations
(\ref{eqn:viscovarpol})  and (\ref{eqn:almpol}) through (\ref{eqn:vispolalm}),
the visibility covariances can be shown to be
\begin{eqnarray}
{\cal V}^{\pm\pm}_{12}
&=&\sum_\ell(C_\ell^E+C_\ell^B)W_\ell^{\pm\pm},\\
{\cal V}^{\pm\mp}_{12}
&=&\sum_\ell(C_\ell^E-C_\ell^B)W_\ell^{\pm\mp},
\end{eqnarray}
where
\beq
W_\ell^{\pm\bm\pm}=\sum_m F_{\mp 2,\ell m}({\bf u}_1)
F_{\bm\mp 2,\ell m}^*({\bf u}_2).
\eeq
In the case where the two observations have different pointing
centers, we once again transform to a coordinate system with
both pointing centers on the equator, separated by an angle $\beta$.
Because the spin-weighted spherical harmonics have azimuthal
dependence $e^{i m\phi}$, the only change is an additional
factor of $e^{im\beta}$:
\beq
W_\ell^{\pm\bm\pm}=\sum_m e^{im\beta}F_{\mp 2,\ell m}({\bf u}_1)
F_{\bm\mp 2,\ell m}^*({\bf u}_2).
\eeq

In order to calculate the correlation between a pair of observations
with arbitrary pointing centers, we simply rotate to
a new coordinate system that places both centers on the equator
before applying the above results.  In performing this rotation,
it is important to remember to transform $P_\pm$ (and hence
$V_\pm$) by $e^{\pm 2i\delta}$ where $\delta$ is the angle
through which the polarization basis directions are rotated by
the transformation.  To be specific, if the change
of coordinates results in a rotation of the $\hat{\bm\theta},\hat{\bm\phi}$
directions at each of the pointing centers by $\delta_1,\delta_2$,
then ${\cal V}_{12}^{\pm\bm\pm}\to{\cal V}_{12}^{\pm\bm\pm}e^{2i(\pm\delta_1
\bm\mp\delta_2)}$.  See the Appendix for an explicit recipe
for finding these angles.

\subsection{Connecting flat-sky to spherical.}

As in the case of temperature anisotropy, we can see the connection
between the spherical and flat-sky calculations of polarization by
considering observations that lie near the equator of our
spherical coordinate system and approximating the sphere by
a cylinder.  By applying the spin-raising operator
\citep[e.g.,][]{LCT} to the
plane-wave approximation to the spherical harmonics (\ref{eq:ylmplane}),
one can show that in this limit
\beq
{}_2Y_{lm}(\rhat)=N_{lm}^{(2)}e^{im\phi}\cases{
\cos (n_{lm}z+\delta_{lm}) & if $l-m$ even\cr
-i \sin (n_{lm}z+\delta_{lm}) & if $l-m$ odd},
\eeq
with
\beq
N_{lm}^{(2)}=N_{lm}l(l+1)\sqrt{(l-2)!\over (l+2)!}=
N_{lm}\sqrt{l(l+1)\over (l+2)(l-1)}
\eeq
and
\beq
\delta_{lm}=2\cos^{-1}\left(m\over l(l+1)\right).
\eeq

By reasoning similar to the previous section we can use this to connect
the spherical harmonic formalism to the flat-sky limit.

\subsection{Example}

Consider a $10\times 10$ mosaic of pointings of an interferometer,
with the same parameters as in the example of Section \ref{sec:cylinder}:
the Gaussian beam width is $\sigma=5^\circ$, and the pointings are 
centered on the equator and separated
by $5^\circ$ in both $\theta$ and $\phi$.  We consider only one baseline
per pointing, with ${\bf u}=22\hat{\bm\phi}$.  Assume that both visibilities
$V_Q$ and $V_U$ are measured (either directly or by measuring 
$V_\pm$).

For any pair of pointings we can define $E$ and $B$ window functions
such that 
\beq
\langle V_{Qi}V_{Qj}^*\rangle=
\sum_\ell (W_\ell^{E}C_\ell^{EE}+W_\ell^{B}C_{\ell}^{BB})
\eeq
where $V_{Qi}$ is the visibility for Stokes $Q$ corresponding
to baseline $i$.  In the limit of infinitely sharp $\ell$-space resolution, 
we would expect $W_\ell^B$ to vanish, since the polarization
would be parallel to the baseline ${\bf u}$.  Inevitably, however,
when only part of the sky is covered (leading to imperfect
Fourier space resolution), there is some mixing 
of $E$ and $B$ modes \citep{LCT,Bunn,BZTO}.

For the case of the two visibilities at the upper corners of the grid,
these window functions are shown in Fig.~\ref{fig:polwindows}.  As in
the scalar case, the correlations are strongly affected by the
rotation of the coordinate basis.  If we incorrectly model the entire
mosaic as flat, treating the basis vectors
$\hat{\bm\theta},\hat{\bm\phi}$ at each point to be parallel, the
correlation between these two pointings would be dramatically
overestimated.  In fact, by treating the sky as flat, we would 
be making two separate errors: treating the two baseline 
vectors as parallel (just as in the case of temperature anisotropy
in Sec.~\ref{sec:cylinder}) and failing to apply the appropriate
transformation to the Stokes parameters $(Q,U)$.

\begin{figure}
\begin{center}
\resizebox{!}{2.5in}{\includegraphics{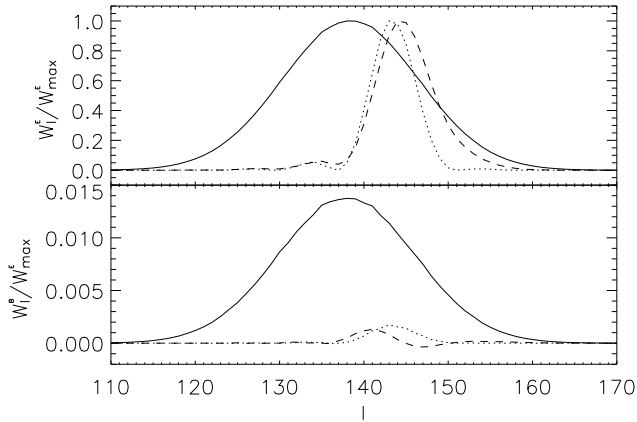}}
\end{center}
\caption{Window functions for a single pointing (solid), the sum of
all pointings (dashed), and the sum of all pointings neglecting sky
curvature (dotted).
\vskip 0.2in}
\label{fig:polmosaic}
\end{figure}

Fig.~\ref{fig:polmosaic} illustrates the improvement of resolution due
to mosaicking in this example.  Like Fig.~\ref{fig:mosaicres}, this
figure shows the autocorrelation window function for a single pointing
as well as that of the sum of all 100 pointings in the grid.  If we neglect
sky curvature, we overestimate the correlation between distant baselines
and hence also overestimate the improvement in $\ell$-space resolution.

\section{Conclusions} \label{sec:conclusions}

Interferometers have been used to great effect in measuring CMB temperature
and polarization anisotropies.  The formalism for analyzing interferometer
data, however, has only been fully developed in the small-field-of-view or
flat-sky limit.  Future experiments which aim for exquisite $\ell$-space
resolution will need to survey large areas of sky -- outside the realm of
validity of the existing formalism.

In this paper we have extended the formalism to the situation where we
can approximate the sky as flat for each individual pointing of the
instrument, but we relax the assumption that the angle between pointings
is also small.  We have connected the full-sky spherical harmonic approach
to the flat-sky Fourier approach in two distinct ways and derived
approximations for the visibility covariance matrix in each.
We find that the cylindrical method of \S\ref{sec:cylinder} and
\S\ref{sec:polcylinder} works in all cases better than the harmonic method
of \S\ref{sec:harmonic} and provides accurate approximations to the full-sky
expressions for individual pointings smaller than $20^\circ$ FWHM.
Mosaicking together many pointings increases the $\ell$-space resolution, but
in the cases considered here the improvement is 
less than would be predicted from the flat-sky formalism, in large part
due to the effects of baseline rotation.
If we neglect sky curvature we overestimate the correlation between distant
baselines and hence also overestimate the improvement in $\ell$-space
resolution.

\acknowledgements

EFB is supported by NSF Grant 0507395 and a Cottrell Award from the
Research Corporation.  EFB thanks the physics departments of Brown
University and MIT for their hospitality during the completion of this
work.  MW is supported by NASA.  We thank Asantha Cooray and Manoj Kaplinghat, 
the organizers of the March
2006 workshop on Fundamental Physics with Cosmic Microwave Background
Radiation, where this work was initiated.

\appendix

\def\theequation{\hbox{A\arabic{equation}}}

In calculating the covariance between visibilities at two different
pointing centers $\rhat_1,\rhat_2$, we must transform to a coordinate
system that places both pointing centers on the equator.  This affects
both the components of the baseline vectors ${\bf u}_1,{\bf u}_2$ and
(in the case of polarization) the Stokes parameters $Q,U$.  We present
here an explicit recipe for performing this transformation.

Throughout this appendix, unprimed symbols will refer to the original
coordinate system, and primed symbols will refer to a coordinate
system $(x',y',z')$ such that both
$\rhat_1$ lies on the $x'$ axis and $\rhat_2$ is in the $x'y'$ plane.
First, choose the $z'$ axis to be perpendicular to both vectors:
\beq
\hat{\bf z}'=\frac{\rhat_1\times\rhat_2}{|\rhat_1\times\rhat_2|}
\eeq
Next, choose the $y'$ axis to be perpendicular to both $\hat{\bf z}'$
and $\rhat_1$:
\beq
\hat{\bf y}'=\rhat_1\times\hat{\bf z}'.
\eeq
Finally, choose $\hat{\bf x}'=\hat{\bf y}'\times\hat{\bf z}'$.
In spherical coordinates $(\theta',\phi')$
defined with respect to the primed
coordinate system, we have $\rhat_1=(\pi/2,0)$ and $\rhat_2=(\pi/2,\beta)$
with $\beta$ such that
$(\cos\beta,\sin\beta)=(\rhat_2\cdot\hat{\bf x}',\rhat_2\cdot\hat{\bf y}')$.

Say that the baseline vector ${\bf u}_i$ ($i=1,2$)
is expressed in the original
(unrotated) spherical coordinate system as
\beq
{\bf u}_i=u_{i\theta}\hat{\bm\theta}+u_{i\phi}\hat{\bm{\phi}}.
\eeq
We need to know the corresponding components in the rotated coordinate
system.  The components of the basis vectors $\hat{\bm\theta},\hat{\bm\phi}$ 
in the
rotated coordinate system are
\begin{eqnarray}
\hat{\bm\theta}&=&(\cos\delta_i)\hat{\bm\theta'}-(\sin\delta_i)\hat{\bm\phi'},\\
\hat{\bm\phi}&=&(\sin\delta_i)\hat{\bm\theta'}+(\cos\delta_i)\hat{\bm\phi'}.
\end{eqnarray}
The easiest way to find the rotation angle $\delta_i$ is to compute
the components of
$\hat{\bm\phi}=(\hat{\bf z}\times\rhat_i)/|\hat{\bf z}\times\rhat_i|$,
and take the dot product $\sin\delta_i=\hat{\bm\phi}\cdot\hat{\bm\theta'}
=\hat{\bm\phi}\cdot(-\hat{\bf z}')$ since $\rhat_i$ is on the equator
in the primed coordinate system.

Once the rotation angles $\delta_1,\delta_2$ are known, the
components of the baseline vectors are
\beq
\left(\matrix{u_{i\theta'}\cr u_{i\phi'}}\right)
=
\left(\matrix{
\cos\delta_i & \sin\delta_i\cr
-\sin\delta_i & \cos\delta_i
}\right)
\cdot
\left(\matrix{u_{i\theta}\cr u_{i\phi}}\right).
\eeq
In calculating the polarization visibilities we replace $(Q,U)$ 
with
\beq
\left(\matrix{Q'\cr U'}\right)
=
\left(\matrix{
\cos 2\delta_i & \sin 2\delta_i\cr
-\sin 2\delta_i & \cos 2\delta_i
}\right)
\cdot
\left(\matrix{Q\cr U}\right).
\eeq

\bibliographystyle{apj}
\end{document}